\documentclass[%
 reprint,
 amsmath,amssymb,
 aps,
 prl,
]{revtex4-1}

\usepackage{color}
\usepackage[version=3]{mhchem} 
\usepackage{gensymb}
\usepackage{amssymb}
\usepackage{upgreek}
\usepackage[usenames,dvipsnames]{xcolor}

\begin{document}

\title{Charge transition levels of quantum emitters in hexagonal boron nitride}
\author{Zai-Quan Xu} 
\email{zaiquan.xu@uts.edu.au}
\affiliation{School of Mathematical and Physical Sciences$,$ University of Technology$,$ Sydney$,$P.O. Box 123$,$ Broadway$,$ New South Wales 2007$,$ Australia.}
\author{Noah Mendelson}
\affiliation{School of Mathematical and Physical Sciences$,$ University of Technology$,$ Sydney$,$P.O. Box 123$,$ Broadway$,$ New South Wales 2007$,$ Australia.}
\author{John A. Scott}
\affiliation{School of Mathematical and Physical Sciences$,$ University of Technology$,$ Sydney$,$P.O. Box 123$,$ Broadway$,$ New South Wales 2007$,$ Australia.}
\author{Chi Li}
\affiliation{School of Mathematical and Physical Sciences$,$ University of Technology$,$ Sydney$,$P.O. Box 123$,$ Broadway$,$ New South Wales 2007$,$ Australia.}
\author{Igor Aharonovich} 
\email{igor.aharonovich@uts.edu.au}
\affiliation{School of Mathematical and Physical Sciences$,$ University of Technology$,$ Sydney$,$P.O. Box 123$,$ Broadway$,$ New South Wales 2007$,$ Australia.}
\author{Milos Toth} 
\email{milos.toth@uts.edu.au}
\affiliation{School of Mathematical and Physical Sciences$,$ University of Technology$,$ Sydney$,$P.O. Box 123$,$ Broadway$,$ New South Wales 2007$,$ Australia.}

\date{\today}

\begin{abstract}
Quantum emitters in layered materials are promising candidates for applications in nanophotonics. Here we present a technique based on charge transfer to graphene for measuring the charge transition levels ($\rm E_t$) of fluorescent defects in a wide bandgap 2D material, and apply it to quantum emitters in hexagonal boron nitride (hBN). Our results will aid in identifying the atomic structures of quantum emitters in hBN, as well as practical applications since $\rm E_t$ determines defect charge states and plays a key role in photodynamics.

\end{abstract}

                              
\maketitle
Van der Waals materials have recently emerged as promising hosts of single photon emitters \cite{rev19}. In particular, hexagonal boron nitride (hBN) is attracting significant attention because it hosts a range of deep trap defects that act as visible and near-infrared quantum emitters with favorable properties such as high brightness, room temperature operation, high chemical stability, and long-term photostability \cite{tran2016quantum, kianinia2018all,exarhos2017optical, shotan2016photoinduced, martinez2016efficient}.  However, little is known about the electronic structure and excitation/relaxation pathways of these emitters. So far, identification of the atomic structures has been hindered by the fact that the emission wavelengths span a broad spectral range exceeding 250~nm \cite{jungwirth2016temperature}.

\begin{figure}[b!]
\resizebox{\columnwidth}{!}{\includegraphics{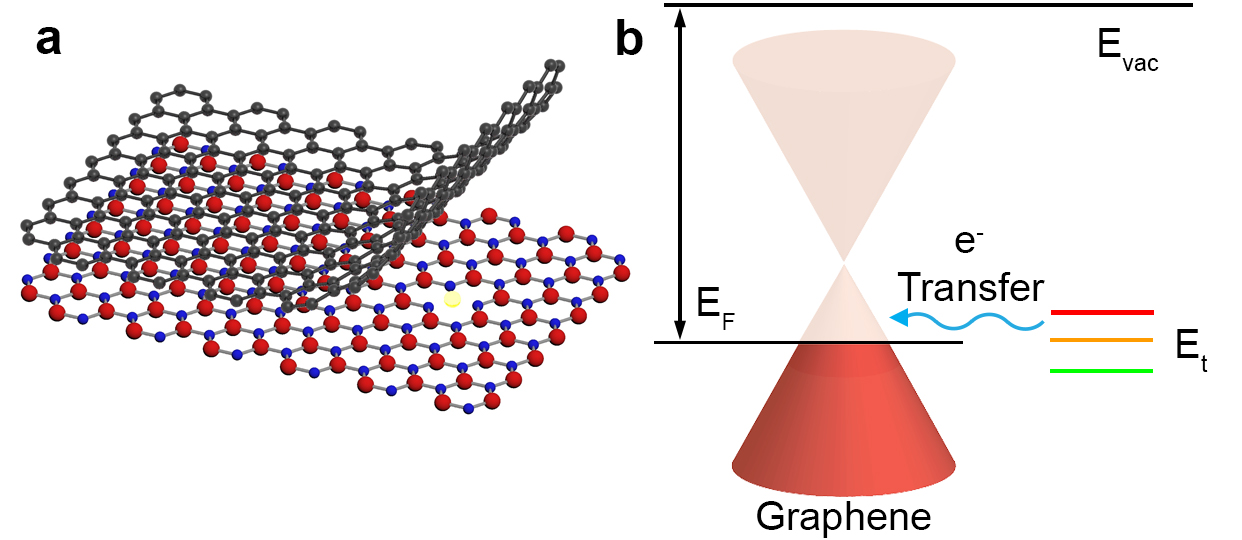}}
\caption{(color online). (a) Schematic illustration of a graphene-hBN heterostructure. (b) Simplified flat-band electron energy diagram showing the charge transition levels of three quantum emitters in hBN $\rm(E_t)$, the Fermi level of graphene $\rm(E_F)$ and the vacuum level $\rm(E_{vac})$. An emitter is ionized via electron transfer to graphene if its charge transition level lies above $E_F$.}
\label{f1}
\end{figure}

A key advantage of hBN is its 2D nature, which enables facile assembly of 2D heterostructures that can be used to manipulate and study charge dynamics \cite{ liu2019room,  jin2017interlayer,wong2015characterization,mendelson2019engineering}. Building upon this idea, we used a graphene-hBN heterostructure, shown schematically in Fig. \ref{f1}(a), to quench quantum emitters in hBN via a non-radiative electron transfer process \cite{tielrooij2015electrical, schädler2019electrical}. Such quenching is expected to occur only if the emitter charge transition level $\rm (E_t)$ lies above the Fermi energy $\rm(E_F)$ of graphene, as is shown in Fig. \ref{f1}(b). We observe it only if the wavelength of the emitter zero phonon line (ZPL) is greater than $\rm\sim 600~nm$, which we attribute to a defect for which $\rm E_t$ is aligned with $\rm E_F$. To confirm this interpretation, we functionalized the graphene overlayer with N-Methyl-2-pyrrolidone (NMP) in order to increase $\rm E_F$ and suppress the quenching of quantum emitters in hBN. Finally, we measured $\rm E_F$ relative to the vacuum level $\rm (E_{vac})$ for both as-transferred and functionalized graphene using environmental photoelectron yield spectroscopy (EPYS). EPYS spectra confirm the shift in $\rm E_F$ expected from NMP functionalization and allow us to quantify the locations of defect charge transition levels within the bandgap of hBN.

Our results are significant because charge transition levels affect defect charge states, as well as the blinking and photo-stability characteristics of quantum emitters -- particularly when they are located near the surface of the host material, as is the case for monolayer and few-layer van der Waals materials \cite{xu2018single,li2017nonmagnetic}. {\color{black} Knowledge of charge transition levels will aid future efforts aimed at identifying the atomic structures of emitters in hBN, as well as applications in which hBN is encapsulated in device structures, or exposed to air or liquids, as is often the case in integrated nanophotonics \cite{schädler2019electrical}, sensing  \cite{Traneaav9180} and fluorescence imaging applications \cite{comtet2019wide,kianinia2018all}.}

\begin{figure}[h!]
\resizebox{\columnwidth*5/6}{!}{\includegraphics{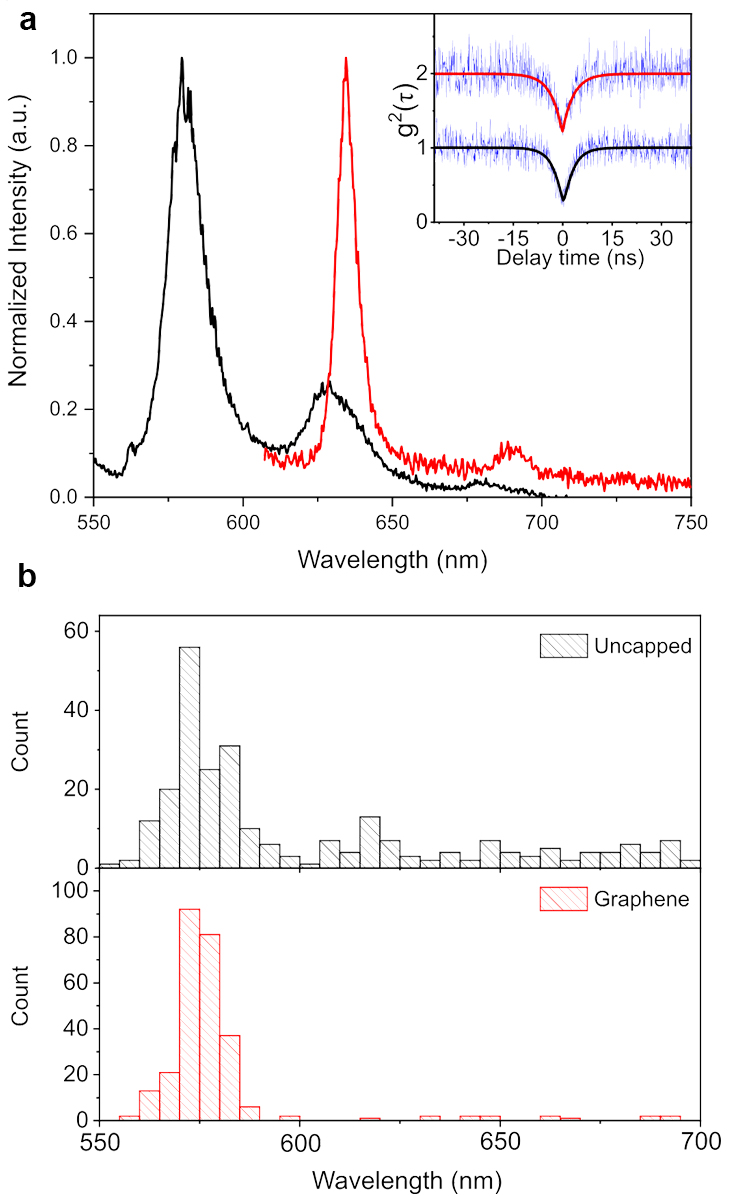}}
\caption{(color online). (a) Photoluminescence spectra of two representative quantum emitters in hBN. Each spectrum consists of a zero phonon line (ZPL) and a phonon sideband (PSB). The inset shows corresponding autocorrelation functions, $\rm g^2(\tau)$, that prove the quantum nature of the emitters. (b) ZPL wavelength histograms obtained from hBN (top) and a graphene-hBN heterostructure (bottom). Graphene causes quenching of most emitters with ZPLs at wavelengths greater than $\rm\sim 600~nm$.}
\label{f2}
\end{figure}

 Commercial hBN was grown on copper by a chemical vapor deposition (CVD) method. AFM analysis revealed a hBN thickness of $\rm(\sim 11.8~nm)$ [Figure S1]. The hBN film and a graphene capping layer were assembled onto an \ce{SiO2} substrate using a PMMA-assisted wet transfer method \cite{tran2017deterministic, wang2019peeling, feng2018imaging}, after which the assembled heterostructure was annealed to remove contaminants and ensure a good contact between layers. Further details of the sample fabrication and characterization are provided in the Supporting Information.

Fig.\ref{f2}(a) shows representative photoluminescence (PL) spectra from two quantum emitters in  hBN. Both SPEs display sharp zero-phonon lines (ZPLs) and a phonon sideband (PSB) separated by an energy detuning of $\rm \sim 140~meV$, which is consistent with previous reports \cite{ tran2016robust}. The ZPL central wavelengths are 579~nm and 633~nm, respectively. Second order autocorrelation functions, $\rm g^2(\tau)$, were measured using a Hanbury-Brown-Twiss interferometer, and are shown in the inset of Fig. \ref{f2}(a) for each emitter. In each case, $\rm g^2(0)<0.5$, confirming the quantum nature of the emissions. 

\begin{figure}[h!]
\resizebox{\columnwidth*5/6}{!}{\includegraphics{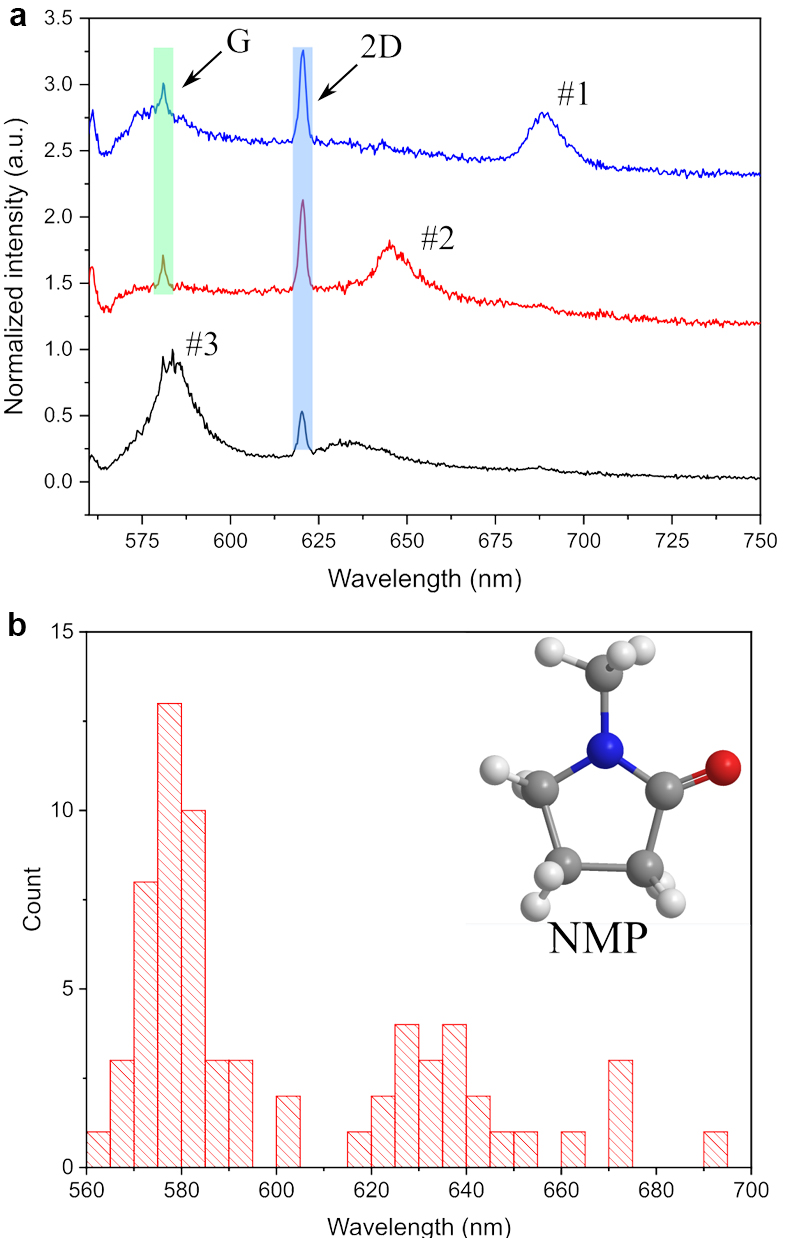}}
\caption{(color online). (a) PL spectra of a graphene-hBN heterostructure functionalized with NMP. The labels highlight ZPLs of three quantum emitters $(\#1, \#2, \#3)$ in hBN, and the G and 2D Raman bands of graphene. (b) ZPL wavelength histogram obtained from the NMP-functionalized graphene-hBN heterostructure. The inset shows a schematic illustration of the molecular structure of NMP.}
\label{f3}
\end{figure}

A histogram of the number of ZPLs as a function of wavelength is shown in the top panel of Fig. \ref{f2}(b) for a pristine hBN film. The histogram has a maximum at $\rm\sim 585~nm$, and spans $\rm\sim 150~nm$, ranging from 550 to 700~nm. Such a broad range of ZPL wavelengths is common in hBN and has been used to argue that a range of defects is responsible for quantum emissions from hBN \cite{jungwirth2016temperature}. The bottom panel of Fig. \ref{f2}(b) shows the same histogram acquired from a heterostructure comprised of graphene and  hBN. PL spectra from the heterostructure consist of emitter ZPLs and PSBs, and Raman peaks that confirm the presence of the graphene overlayer (see the {\color{black}Supporting Information}, and Fig. \ref{f3}(a)). The two histograms are similar at wavelengths shorter than ~600 nm, but the latter is strongly attenuated at wavelengths beyond $\rm\sim 600~nm$. That is, most defects emitting at wavelengths greater than $\rm\sim 600~nm$ are quenched selectively by the graphene overlayer (we note that a minority of ZPLs in this spectral range are not quenched, likely due to defects/pinholes in the graphene and residuals at the graphene-hBN interface). 

The graphene overlayer also reduces the luminescence efficiency of all emissions from the hraphene-hBN heterostructure, {\color{black} see Supporting Information}. We attribute this to non-radiative energy transfer (NRET) to graphene, a well known phenomenon \cite{gaudreau2013universal, scavuzzo2019electrically} that occurs in conjunction with the electron transfer effect studied in the present work. 

\begin{figure}[h!]
\resizebox{\columnwidth*5/6}{!}{\includegraphics{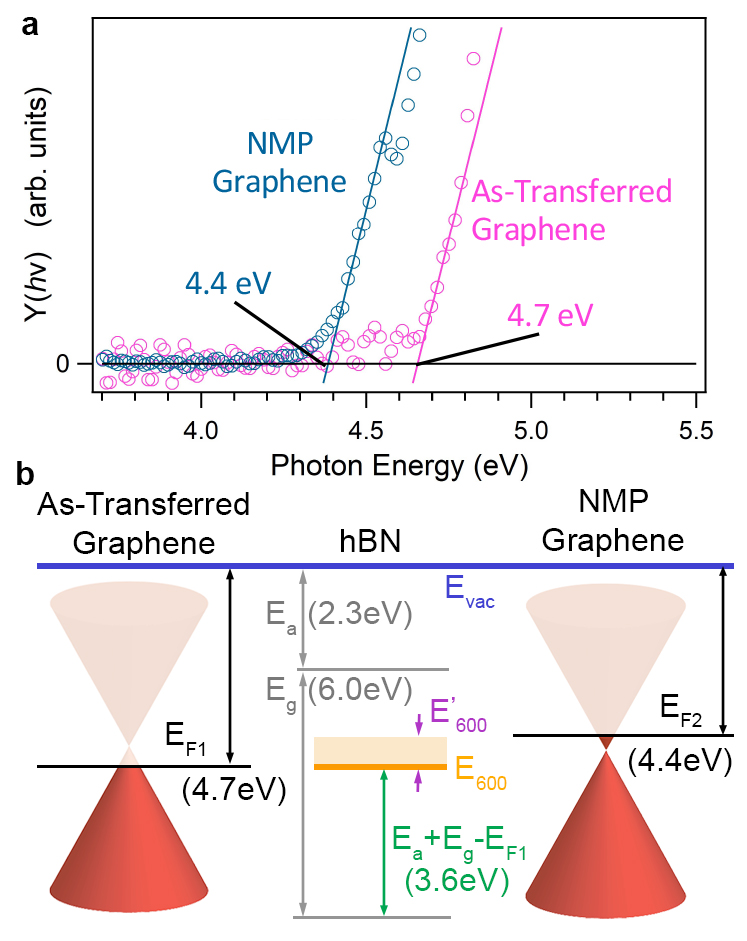}}
\caption{(color online). (a) Photoelectron yield spectra showing the Fermi energy of as-transferred graphene $(\rm E_{F1}=4.7~eV)$ and NMP-functionalized graphene $(\rm E_{F2}=4.4~eV)$. (b) Simplified flat-band electron energy diagram showing the vacuum level $\rm(E_{vac})$, $\rm E_{F1}$, $\rm E_{F2}$, hBN bandgap $\rm (E_g)$, hBN electron affinity $\rm E_a$, and the charge transition level of a quantum emitter with a ZPL at 600 nm $\rm (E_{600})$, which lies $\rm E_a+E_g-E_{F1}$ electronvolts above the valance band maximum. $\rm E^{'}_{600}$ denotes an energy range of $\rm\sim 0.3~eV$ which contains the charge transition levels of emitters with ZPL wavelengths greater than 600~nm.
}
\label{f4}
\end{figure}

We attribute the selective quenching (i.e., SPEs above 600~nm) seen in Fig. \ref{f2} to electron transfer from emitters in hBN to graphene. Electron transfer is expected only if the emitter charge transition level is located above the Fermi level of graphene, as is illustrated in Fig. \ref{f1}(b). To confirm this interpretation, we functionalized graphene with NMP {\color{black}(Fig. S1)}, which is known to act as an electron donor to graphene. NMP is therefore expected to raise $\rm E_F$ towards the vacuum level, and inhibit the electron transfer from defects for which the elevated $\rm E_F$ is located above the defect charge transition level. Fig. \ref{f3}(a) shows PL spectra of three emitters in hBN with an overlayer of NMP-functionalized graphene. The spectra contain three ZPLs centred on approximately 583, 645 and 688~nm, as well as the G and 2D Raman bands of graphene. Critically, the Raman peaks indicate that graphene is present on the sample after the NMP functionalization step. A histogram of the ZPL wavelengths, shown in Fig. \ref{f3}(b), reveals that the quenching of quantum emitters in hBN is inhibited by the presence of NMP. This is consistent with the proposed electron transfer mechanism since NMP is expected to raise the Fermi level of graphene towards the vacuum level. The histogram reveals that the quenching is absent at wavelengths of up to $\rm\sim 700~nm$, implying that the shift in $\rm E_F$ caused by NMP must be greater than or equal to $\rm\sim 0.3~eV$. To confirm this, we used EPYS to measure $\rm E_F$ of as-transferred and NMP-functionalized graphene in Ar and NMP vapor environments. In the EPYS technique, the photoelectron emission yield is measured in a gaseous environment as a function of incident photon energy, and the phototelectron emission threshold is the ionization energy of the sample \cite{shanley2014localized} (i.e., the Fermi energy in the case of graphene). EPYS spectra from as-transferred and NMP-functionalized graphene are shown in Fig. \ref{f4}(a) {\color{black}(a detailed description of the measurement is provided in the Supporting Information)}. The photoelectron emission threshold is 4.7 and 4.4~eV for as-transferred and NMP-functionalized graphene, respectively. The value of 4.7~eV lies 0.2~eV below the Dirac point of charge nuetral graphene (4.5~eV) \cite{xu2012direct}, confirming the expected initial p-type doping of graphene upon polymer transfer to SiO$_2$. NMP functionalization shifted the graphene Fermi level towards the vaccum level by $\rm\sim 0.3~eV$, resulting in a slightly n-type graphene overlayer. The measurement of a 0.3~eV upward shift in the Fermi level caused by NMP is consistent with the ZPL histograms shown in  Fig. \ref{f2}(b) and \ref{f3}(b).

We further confirmed the change in graphene doping level by Raman spectroscopy of as-transferred and NMP-functionalized graphene. The intensity ratio of the G and 2D modes as well as broadening and shifts of the G peak provide a sensitive measure of the relative doping level of graphene \cite{das2008monitoring}. Our spectra {\color{black}(shown in the Supporting Information)} confirm that the as-transferred graphene is p-type, and that the Fermi level shifts to just above the Dirac point, yielding slightly n-type characteristics upon NMP functionalization). The Raman data therefore serves as independent complimentary evidence for our interpretation of the EPYS spectra and the ZPL histograms.

 {\color{black} Fig. \ref{f4}(b) summarizes our results on a simplified electron energy diagram. It  shows $\rm E_{vac}$, the Fermi level of as-transferred and NMP-functionalized graphene ($\rm E_{F1}$ and $\rm E_{F2}$, respectively), the band gap of hBN {\color{black} $\rm(E_g=6.0~eV)$} \cite{ cassabois2016hexagonal}, and the charge transition level of a quantum emitter with a ZPL at 600~nm $\rm(E_{600})$. The figure also shows the electron affinity of hBN {\color{black}$\rm (E_a \sim 2.3~eV)$} \cite{choi2017gas}. $\rm E_{600}$ is approximately aligned with $\rm E_{F1}$, which places it (i.e., the charge transition level of an emitter with a ZPL of 600~nm) approximately {\color{black}3.6~eV} above the valance band maximum of hBN.  
 
 The charge transition levels of emitters with ZPLs beyond 600~nm lie between $\rm E_{F1}$ and $\rm E_{F2}$, an energy range of $\rm\sim 0.3~eV$, indicated by $\rm E^{'}_{600}$ in Fig. \ref{f4}(b)---these are the emissions that are quenched by graphene, and restored by NMP. For emitters below 600~nm, the charge transition levels lie below $\rm E_{F1}$---these emissions are not quenched by graphene.

In summary, we presented a technique for measuring charge transition levels of fluorescent defects within the bandgap of hBN, based on defect ionisation through charge transfer to a graphene overlayer. The precision of the technique is determined by the proximity of the Fermi level of graphene to the charge transition level, and the degree to which $\rm E_F$ can be altered by functionalisation of the graphene.  Knowledge of charge transition levels will aid theoretical modelling studies aimed at identifying the atomic structures of emitters, as well as the development of strategies for mitigating blinking and spectral diffusion effects which can be severe in hBN \cite{stern2019spectrally, li2017nonmagnetic}. Our work also underscores the advantages of SPE sources embedded in atomically thin materials, as the ability to emit light from near-surface defects and to assemble heterostructures offers a degree of freedom not available in 3D materials, shown here by controlling emission through a graphene overlayer that acts as spectral filter. The ability to extract and inject charges into hBN defects on a nm scale also opens up exciting opportunities for electro-luminescent devices, and the electronic readout of the defect states \cite{wong2015characterization, ju2014photoinduced, brenneis2015ultrafast}.}

\section{Acknowledgements}
Z.X. and N.M. contributed equally to this work. The authors thank Jacqueline Echeverria for experimental assistance.  We gratefully acknowledge financial support from the Australian Research Council (DP180100077, DP190101058, LP170100150), the Asian Office of Aerospace Research and Development grant FA2386-17-1-4064, and the Office of Naval Research Global under grant number N62909-18-1-2025.

\bibliographystyle{apsrev4-1}
\bibliography{grapheneREFs}
\end{document}